 \documentclass[aps,pra,twocolumn,lengthcheck,superscriptaddress,showpacs,preprintnumbers,amsmath,amssymb]{revtex4}

 \usepackage{graphicx}
 \usepackage{dcolumn}
 \usepackage{bm}

\begin{document}


 \title{Phase diagram for ultracold bosons in optical lattices and superlattices}

 \author{P. Buonsante}
 \affiliation{Dipartimento di Fisica, Politecnico di Torino and I.N.F.M, Corso Duca degli Abruzzi 24 - I-10129 Torino (ITALIA)}%
 \author{A. Vezzani}
 \affiliation{Dipartimento di Fisica, Universit\`a degli Studi di Parma and I.N.F.M., Parco Area delle Scienze 7/a I-43100 Parma (ITALIA)}%

 \date{\today}

 \begin{abstract}
 We present an analytic description of the finite-temperature phase diagram of the Bose-Hubbard model, successfully describing the physics of cold bosonic atoms trapped in optical lattices and superlattices. 
 Based on a standard statistical mechanics approach, we provide the exact expression for the boundary between the superfluid and the normal fluid by solving  the self-consistency equations involved in the mean-field approximation to the Bose-Hubbard model.
 The zero-temperature limit of such result supplies an analytic expression for the Mott lobes of superlattices, characterized by a critical fractional filling.
 \end{abstract}

 \maketitle

 \section{Introduction}
 \label{S:in}

 The standing wave produced by the interference among counterpropagating laser 
 beams gives rise to a periodic potential commonly used to fragment and trap 
 clouds of ultracold (possibly condensed) alkali atoms 
 \cite{A:BAnderson,A:Orzel,A:Morsch}. 
 The local minima of such trapping potential are the sites of the 
 so-called {\it optical lattice} \cite{A:Jaksch}. 
 The use of multiple wavelength laser beams allows to obtain superlattices,
 namely more structured periodic potentials characterized by a spatial
 modulation of the depth of the lattice wells
 \cite{A:Guidoni97,A:Roth03,A:Blakie,CM:Santos,A:Aizenman}.

 In the case of bosonic atoms cooled to within the lowest Bloch band of the 
 periodic potential, it can be shown \cite{A:Jaksch} that the physics of the 
 system is described by the well-known Bose-Hubbard Hamiltonian \cite{A:Fisher}
 \begin{equation}
 \label{E:BH}
 H \!=\!\sum_j \left[ \frac{U}{2}n_j (n_j\!-\!1)\! -\!(\mu-v_j) n_j
 \!-\!t  a_j \sum_{h\sim j} \tau_{jh} a_h^+ \right]
 \end{equation}
 where the subscripts $j,h$ label the sites of the optical lattice, $a_j^+$  ($a_j$) creates (annihilates) a boson at site $j$, $n_j =
 a_j^+a_j$ counts the bosons at site $j$, and the symbol $\sim$ restricts the sum over $h$ to the nearest neighbours of $j$. 
 This is obtained by a scheme analogous to the tight-binding approximation commonly adopted for the study of electrons in solids, i.e. by expanding the state of the system onto a set of  wavefunctions localized at the local minima of the trapping potential. 
 The parameters $U$, $t$, $\tau_{jh}$ and $v_j$ are hence given in terms of 
 overlap integrals between the localized wavefunctions at neighbouring sites 
 and the trapping potential. Specifically,
 $U$ represents the repulsive boson-boson interaction, $v_j$ is the local potential 
 at site $j$ and $t\cdot\tau_{jh}$ is the 
 hopping amplitude between sites $j$ and $h$. More in detail, $t$ is a global
 scaling factor  determined by the laser 
 intensity, while $\tau_{jh}=\tau_{hj}$ is a local parameter depending on the
 details of the optical potential in the region between lattice sites $j$ and
 $h$ (see Fig.~\ref{F:SLpotper} for a schematic representation of the optical
 superlattices considered in the following).
 A fine tuning of 
 parameters $v_j$, $t$ and $\tau_{ij}$ can be in principle obtained via 
 suitable variations of experimental parameters such as the intensity, the 
 frequency and the geometric setup of the laser beams producing the optical
 (super)lattice \cite{N:Feshback}. 
 For generic superlattices $\tau_{jh}$ and $v_j$ 
 are periodic functions of the lattice labels
 , whereas in the case of 
 the usual single-wavelength optical lattices they are usually assumed to be
 site-independent, so that one
 can set $\tau_{jh}=1$ and $v_j=0$ without loss of generality.
 %
 %

 The parameter $\mu$ appearing in Hamiltonian (\ref{E:BH}) is the usual chemical potential of the grand canonical statistical approach, and is fixed by the total number of bosons in the system.

 Hamiltonian (\ref{E:BH}) is also strictly related to systems other than the one under concern, such as Josephson junction arrays and quantum spin systems on lattices \cite{A:FazioPR,CM:Garcia}. The hallmark of such class of systems is no doubt the quantum phase transition between a superfluid and a (Mott) insulator phase \cite{A:Fisher} originating from the competition between the repulsive and  kinetic term of the Hamiltonian, whose magnitude are proportional to the parameters $U$ and $t$, respectively. 
 The fine tuning of these parameters made possible by the striking progress in
 optical lattice techniques  allowed Greiner and co-workers to observe the
 superfluid-insulator transition in a recent breakthrough experiment \cite{A:Greiner}.
 More in general, ultracold neutral atoms make an ideal benchmark for testing the properties of widespread models of condensed matter physics \cite{A:Zwerger}.

 It is worth recalling that the above quantum phase transition is rigorously
 present only at zero temperature \cite{B:Sachdev}, whereas at finite
 temperature thermal fluctuations induce a classical phase transition between a
 superfluid and a normal phase. However, at sufficiently low temperatures, a
 remnant of the insulating phase still persists within the normal phase. 
 Indeed in these conditions it is possible to observe a sharp crossover between a compressible normal fluid and a phase characterized by a vanishing compressibility, which, for all practical purposes, can be considered a Mott insulator \cite{A:Sheshadri,A:Dicker}. 

 The zero-temperature phase diagram of the Bose-Hubbard model has been widely studied using a variety of techniques, including the mean-field approaches \cite{A:Sheshadri,A:Amico,A:vanOosten,CM:Jain}, strong coupling perturbative expansion \cite{A:Freericks1,A:Elstner99a,CM:scpe}, density matrix renormalization group \cite{A:Kuehner} and of course Quantum Monte Carlo simulations \cite{A:Batrouni,A:Kashurnikov96}. 

As to the finite-temperature case, some early numerical results concerning the homogeneous lattice are reported in Ref.~\onlinecite{A:Kampf}, where a {\it coarse graining} mean-field approach is adopted, and in 
Ref.~\onlinecite{A:Sheshadri}, based on a random phase approximation
refining the mean-field approach therein proposed.  In Ref.~\onlinecite{A:Amico96}
the weak-repulsion limit is addressed, and some  results are obtained
numerically within a different mean-field scheme, based on the linearization
of the repulsive term rather than on the decoupling of the hopping term, like
in Ref.~\onlinecite{A:Sheshadri}.
 Quite recently, Dickerscheid and co-workers
\cite{A:Dicker} adopted a slave-boson technique allowing to include the finite
temperature effects in the mean-field picture of Ref.~\onlinecite{A:Sheshadri}. The ongoing interest in the
issue under examination is
further confirmed by Ref.~\onlinecite{CM:Giampaolo} ---where a multiband model
is addressed --- and Ref.~\onlinecite{CM:Plimak} --- where some analytic
results are obtained by interpolating two different perturbative schemes and
subsequently checked against density matrix renormalization group simulations.
All of the above listed results have been obtained either numerically
or applying some further approximation such as
 introducing  tight restrictions on the  number of particles per site.
 Note however that, despite such
 restrictions, the latter approach may prove sufficient to give satisfactory
 results within circumscribed regions of the phase diagram.

 Here we focus on the mean-field approach to Hamiltonian (\ref{E:BH}) proposed by Sheshadri {\it et al}.
 \cite{A:Sheshadri}, and, for any temperature $T$, we determine analytically
 the boundary of the superfluid domain of the phase diagram thereof.
In this framework the phases of the system are characterized in terms of the
 so-called {\it superfluid order parameter}, to be determined as the stable 
fixed point of a self-consistency equation. 
Detailedly, this parameter vanishes in the normal
 fluid phase, whereas it has a finite value in the superfluid phase. 
We determine the critical boundary between these phases  by
discussing the (parameter dependent) stability of the fixed point
 corresponding to the normal phase. 
Furthermore we discuss the above-mentioned crossover between
the compressible normal fluid and the {\it insulator-like} phase taking
place outside the superfluid domain. 
Other than the usual $d$-dimensional homogeneous lattice, we consider
a generic one-dimensional $\ell$-periodic superlattice, providing a solution
in terms of the maximal eigenvalue of a $\ell\times\ell$-matrix.
 Explicit results are given for the
 the 2-periodic and for a special case of the 3-periodic superlattice, 
where such maximal eigenvalue can be easily worked out.
The zero-temperature
 phase-diagram of the above mentioned systems is  recovered taking the
 appropriate limit in our results. In particular, for superlattices,
 we find that rational
filling lobes appear besides the usual integer-filling Mott domains.
Also, we observe that the occurrence of the latter can be 
prevented with  suitable choices of the supercell potential profile.
All of our results prove equivalent to those obtained adopting the
standard numerical algorithm, based on a self-consistent iterative 
procedure. 

The plan of this paper is as follows. In section~\ref{S:MFA} we briefly recall
the mean-field approach presented in Ref.~\onlinecite{A:Sheshadri} and
introduce the finite-temperature self-consistency condition.
In Section~\ref{S:HC} we shortly address the homogeneous lattice case, 
providing the exact expression for the boundary of the superfluid domain, 
and discussing the crossover between the compressible normal fluid
and a insulator-like phase. These results are extended to the case of
periodic superlattices in Sec.~\ref{S:NHC}, where the two above-mentioned
special cases are explicitly considered. Most of the technical details of our
derivation are confined to Appendix \ref{S:det}.
Section \ref{S:C} contains our conclusions.

\section{Mean Field Approximation}
\label{S:MFA}
The mean-field approach to the Bose-Hubbard model introduced by Sheshadri and co-workers
\cite{A:Sheshadri} relies on the standard approximation
\begin{equation}
\label{E:mfa}
a_j a_h^+ \approx \langle a_j \rangle a_h^+ + a_j \langle a_h^+ \rangle 
- \langle a_j  \rangle  \langle a_h^+  \rangle 
\end{equation}
where $\alpha_j \equiv  \langle a_j \rangle  =  \langle a_j^+ \rangle $ is the
so-called {\it superfluid parameter} \cite{N:real}, to be determined self-consistently.
Indeed, equation~(\ref{E:mfa})  allows to recast Hamiltonian (\ref{E:BH}) 
as the sum of terms containing on-site operators only:
\begin{eqnarray}
\label{E:mfH1}
{\cal H}   &=& \sum_{j=1}^M {\cal H}_j \\
\label{E:mfH2}
{\cal H}_j &=& \frac{U}{2} n_j(n_j-1) -(\mu-v_j) n_j\nonumber \\
 & &-t \left(a_j+a_j^+-\alpha_j \right) \sum_{h\sim j} \tau_{jh}\alpha_h
\end{eqnarray}
where $M$ is the number of lattice sites. Note that, unlike Hamiltonian (\ref{E:BH}), the mean-field Hamiltonian $\cal H$ features single boson terms, and therefore it does not conserve the total number of bosons.

A qualitative zero-temperature phase diagram of the BH model 
can be obtained by evaluating the expectation value $\langle \cdot \rangle$ on
the ground state of Hamiltonian~(\ref{E:mfH1}). Such evaluation must be
performed self-consistently, since the ground state of $\cal H$ itself depends
on the set of superfluid parameters  $\{\alpha_j\}$.
In the particular case of homogeneous lattices, translational
invariance yields $\alpha_j = \alpha$, and one is left with ($M$ identical
copies of) a single-site problem.
The resulting phase diagram consists of a superfluid region, where $\alpha~>
~0$, and a series of Mott-insulator lobes, where $\alpha=0$ and the local
density $\langle n_j \rangle$ is pinned to an integer value (and hence the
system is in an incompressible state, $\partial_\mu \langle n_j \rangle=0$ ).
The boundaries of these Mott lobes have been determined numerically in the
original paper \cite{A:Sheshadri}, while  their analytical
expression has been reported in a quite recent work \cite{A:vanOosten}.

More in general this mean-field approach has been adopted for studying the superfluid-insulator transition in some inhomogeneous situations.
An harmonic confining potential $v_j \propto
(j-j_0)^2$ is considered in Ref.~\onlinecite{A:Polkov}, whereas the effect of
topological inhomogeneity is addressed in Ref.~\onlinecite{A:BHComb}.

Here we are interested in the thermodynamics of the 
system, and we adopt the standard grand-canonical 
statistical mechanic approach, 
\begin{equation}
\label{E:exv}
\langle {\cal O} \rangle = \frac{{\rm Tr}({\cal O} e^{-\beta {\cal H}}
  )}{{\rm Tr}(e^{-\beta {\cal H}})}, 
\end{equation}
where the trace is evaluated on the whole Fock space.
Exploiting the site-decoupling of the mean-field Hamiltonian  ${\cal H}$, Eq.~(\ref{E:mfH1}), the self-consistency conditions become
\begin{eqnarray}
\alpha_j \!\! &=& \!\! \langle a_j \rangle = \frac{{\rm Tr}(a_j e^{-\beta {\cal H}}
  )}{{\rm Tr}(e^{-\beta {\cal H}})} \nonumber \\
\label{E:sce}
\!\!&=& \!\!\frac{ {\rm Tr}(a_j e^{-\beta {\cal H}_j})\prod_{k\neq j} {\rm Tr}( e^{-\beta {\cal H}_k})}{\prod_{k=1}^M {\rm Tr}( e^{-\beta {\cal H}_k})} \!=\! \frac{{\rm Tr}(a_j e^{-\beta {\cal H}_j} )}{{\rm Tr}(e^{-\beta {\cal H}_j})}
\end{eqnarray}
where the traces in the second line are evaluated on the single site Fock space.
Note that the superfluid parameters $\alpha_j$ can be safely assumed to be
real since both ${\cal H}_j$ and $a_j$ are real operators.

In the following sections we illustrate how it is possible to determine
analytically the critical condition for superfluidity, i.e. for the existence
of a stable solution of the self-consistency equations (\ref{E:sce}) with
$\alpha_j \neq 0$.

\section{Homogeneous case}
\label{S:HC}
When $v_j=0$ and $\tau_{j k}=\tau$ the system is translationally invariant and, similar to the
above-recalled zero-temperature case, Eqs.~(\ref{E:sce}) reduce to ($M$ identical
copies of) a single consistency equation. Dropping the site-labeling
subscripts one gets $\alpha_j=\alpha = \langle a \rangle$ and
\begin{equation}
{\cal H}_j = \bar {\cal H} = \frac{U}{2} n(n-1) - \mu n -2 d\,t \alpha (a+a^+)
\end{equation}
where $d$ is the dimension of the lattice and we discarded the constant term
$2td\alpha^2$ since it can be factored out from both the numerator and the
denominator of Eq.~(\ref{E:exv}).

In this simple case the self-consistency constraint, Eq.~(\ref{E:sce}), depends on the (site-independent) superfluid parameter $\alpha$, and it is met when the latter is a stable fixed point. It is easy to check that $\alpha=0$ is a fixed point of Eq.~(\ref{E:sce}) whose stability depends on the parameters
$U$, $\mu$ and $t$. In particular, when $\alpha=0$ is unstable, a stable
solution $\alpha>0$ is expected and the system is in a 
superfluid state.

\begin{figure}
\begin{center}
\includegraphics[width=8.5 cm]{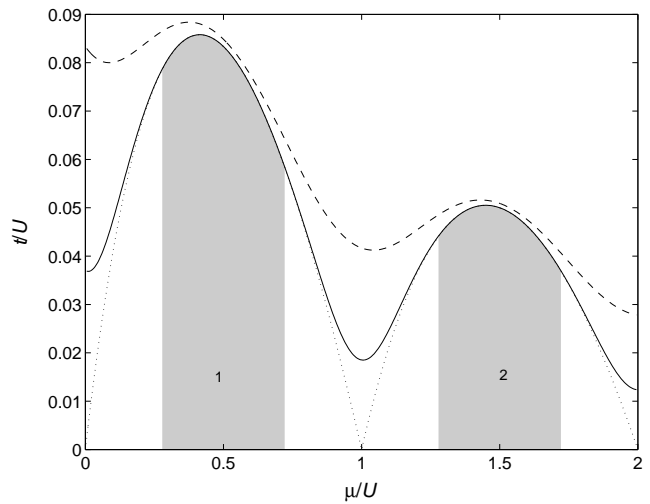}
\caption{\label{F:hom2d} 
Finite temperature mean-field phase diagram for the homogeneous
one-dimensional lattice. The dotted, solid and dashed lines are the critical
curves, Eq. (\ref{E:tc}), for $T/U=0$, $T/U=0.04$ and $T/U=0.1$, respectively.
In every case, the superfluid domain is above the relevant critical curve. For
$T/U=0$, the region below the (dotted) critical curve consists of
two disjoint Mott lobes. For $T/U=0.04$ (solid curve) a sharp crossover
between a normal fluid phase and a insulator-like phase is present. The gray
areas are the insulator-like regions as evaluated setting $\epsilon=10^{-3}$
in Eq.~(\ref{E:int}). As discussed in the text, the particle density is very
close to an integer value (also shown) inside these regions.  For $T/U=0.1$
(dashed curve) no insulator-like regions are present (for the same value of $\epsilon$).}
\end{center}
\end{figure}
\begin{figure*}
\begin{center}
\includegraphics[width=16 cm]{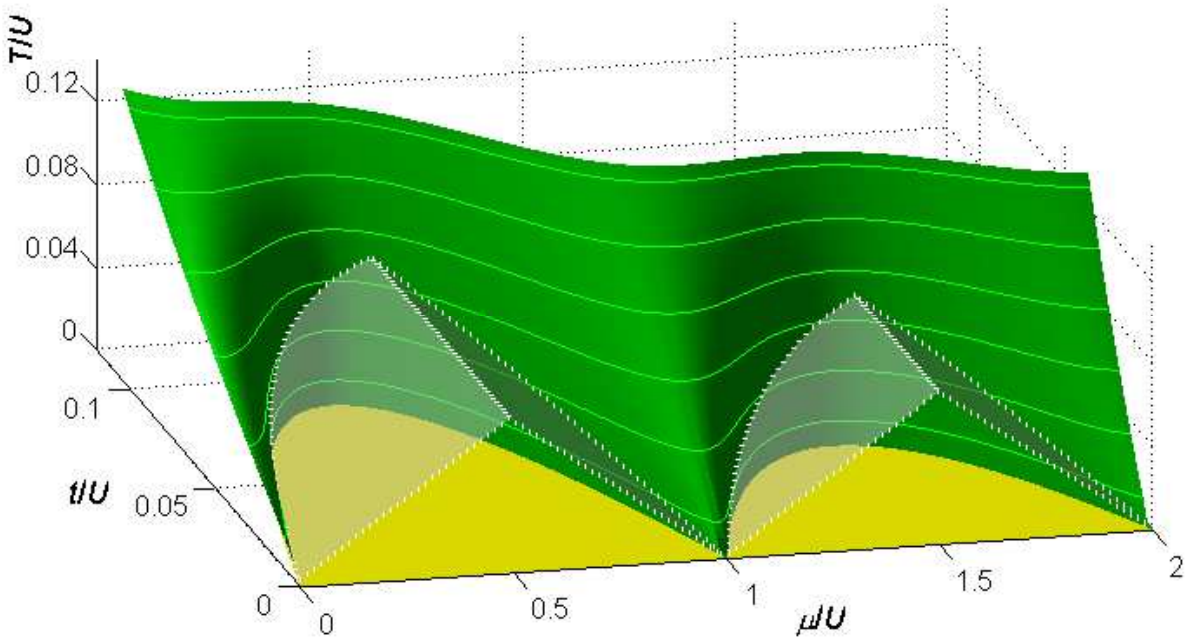}
\caption{\label{F:hom3d}(Color online) Three dimensional representation of the
  finite-temperature mean-field phase diagram for a homogeneous
  one-dimensional lattice. Dark gray (green): critical surface,
  Eq.~(\ref{E:tc}). Light gray (yellow): zero-temperature Mott
  lobes. White-rimmed transparent surfaces: boundaries between the 
  insulator-like (below) and normal fluid (above) regions. Figure \ref{F:hom2d} is
  obtained by cross-sectioning this
  figure at the relevant  values of $T/U$.    }
\end{center}
\end{figure*}

In Appendix \ref{S:det}  we show that the critical curve defining the
superfluid domain border is
\begin{equation}
\label{E:tc}
t_c(U,\mu,\beta) =  \frac{\sum_{k=0}^\infty e^{\beta(\mu k-\frac{U}{2} k(k-1))}}
{2 d \sum_{k=0}^\infty  Q_k(U,\mu) e^{\beta(\mu k-\frac{U}{2} k(k-1))}}
\end{equation}
where 
\begin{equation}
\label{E:lobo}
Q_k(U,\mu) = \frac{\mu+U}{(\mu-U k)[U(k-1)-\mu]}
\end{equation}
Note that when $\alpha=0$ the expectation value of the
particle density does not depend on $t$, being simply
\begin{equation}
\label{E:rho}
\rho(U,\mu,\beta)=\langle n \rangle = 
 \frac{\sum_{k=0}^\infty k e^{\beta(\mu k-\frac{U}{2} k(k-1))}}
{\sum_{k=0}^\infty  e^{\beta(\mu k-\frac{U}{2} k(k-1))}}
\end{equation}
Hence, unlike the zero-temperature case, the $\alpha=0$ region in general does
not yield integer particle density. 
However it can be shown that, for sufficiently 
low temperatures, there are $\mu$ intervals where $\rho(U,\mu,\beta)$ is
practically constant. This happens when  a single term of the sums in
(\ref{E:rho}), whose label we denote $k^*$, outweighs the remaining terms, i.e. when
\begin{equation}
e^{-\frac{\beta U}{2}(k^*- \frac{\mu}{U}-\frac{1}{2})^2} > \epsilon^{-1}
 e^{-\frac{\beta U}{2}(k^*\pm 1- \frac{\mu}{U}-\frac{1}{2})^2}
\end{equation}
where $\epsilon$ is a small parameter. 
The last equation identifies the interval
\begin{equation}
\label{E:int}
k^*-1-\frac{\ln(\epsilon)}{\beta U}< \frac{\mu}{U}< k^* +\frac{\ln(\epsilon)}{\beta U}
\end{equation}
where $\rho(U,\mu,\beta)\approx k^*$. More precisely, expanding
Eq.~(\ref{E:rho}) with respect to $\epsilon$ and considering only the linear
order, one gets $|\rho-k^*|\approx \epsilon$. In the normal fluid region between two subsequent intervals of the form (\ref{E:int}) the particle density can be described considering only the terms $k^*$ and $k^*+1$ in Eq.~(\ref{E:rho})
\[
\rho(U,\mu,\beta) \approx \frac{k^*+(k^*+1)e^{\beta(\mu-U k^*)}}{1+e^{\beta(\mu-U k^*)}}
\]
Last equation clearly shows the crossover between two subsequent insulator-like regions. Of course, when $\beta< -2 \ln(\epsilon)/U$ there is no $\mu$ satisfying the Eq.~(\ref{E:int}), and the plateau-like behaviour of $\rho(U,\mu,\beta)$ disappears.

The same line of reasoning allows to obtain quite
straightforwardly the exact phase
diagram of the model in the zero temperature limit \cite{A:vanOosten}. Indeed,
when $\beta\to\infty$ only the terms labeled by $k=k^*$ survive in Eq.~(\ref{E:tc}), so
that the critical curve in the $\mu/U$-$t/U$ phase diagram is
\begin{equation}
\frac{t_c(U,\mu,\infty)}{U} =
\frac{(\frac{\mu}{U}-k^*)(k^*-1-\frac{\mu}{U})}{2 d(\frac{\mu}{U}+1)}
\end{equation}
and, according to Eq.~(\ref{E:int}), $\mu \in [k^*-1,k^*]$.

The boundary of the superfluid
domain at different temperatures, along with the crossover between the normal
fluid and the Mott insulator (when present) is shown in Figs.~\ref{F:hom2d}
and ~\ref{F:hom3d} for a homogeneous lattice with $d=1$. As it is evident from
Eq.~(\ref{E:tc}), the results for $d>1$ are obtained by a suitable rescaling
of $t_c$.

\section{Superlattices}
\label{S:NHC}
We now turn to the case of superlattices, where the parameters 
appearing in Hamiltonian (\ref{E:mfH1}) are periodic 
functions of the site label. Our approach can be applied to a generic 
$d$-dimensional superlattice, but, for the sake of clarity, 
here we focus on the one-dimensional $\ell$-periodic case, 
$v_j=v_{j+\ell}$ and
$\tau_{j,h}=(\delta_{h,j+1}+\delta_{h,j-1})\tau_{j+\ell,h+\ell}$.  
Note that this choice is not merely dictated by the ensuing notational 
simplification, but also experimentally relevant \cite{A:Morsch,A:Orzel}. 

Since the superfluid parameters mirror the $\ell$-periodicity of the
superlattice, $\alpha_j=\alpha_{j+\ell}$, the
self-consistency conditions~(\ref{E:sce}) reduce to $\ell$ independent
equations. As in the homogeneous case, the choice $\alpha_h=0$  for all $h$'s  
is a fixed point of Eq.~(\ref{E:sce}), and only when it is unstable the system is expected to be in a superfluid state.
According to a standard approach, the stability of such fixed point can be
discussed based on the spectrum of the matrix linearizing the map defined by Eq.~(\ref{E:sce}) in the
vicinity of the configuration $\alpha_h=0$. More in detail, the fixed point is
stable only if the modulus of the maximal eigenvalue of such matrix is 
lower than one.
%
%
By adopting a
calculation technique similar to the one of the homogeneous case (see Appendix \ref{S:det}), the linearized map turns out to be
\begin{equation}
\label{E:matmap}
\alpha_h \approx t
\sum_{h'=1}^{\ell} F_{h,h'} \alpha_{h'}
\end{equation}
where, introducing $\mu_h=\mu-v_h$ and ${\cal T}_{h,h'} =\tau_{h\,h'}
(\delta_{h\,h'+1}+\delta_{h\,h'-1})+\tau_{\ell,\ell+1}( \delta_{h\,1}
\delta_{h'\,\ell} + \delta_{h'\,1} \delta_{h\,\ell})$, 
\begin{equation}
\label{E:matrixMSL}
F_{h,h'} = {\cal T}_{h,h'}
\frac{\sum_k Q_k(U,\mu_h)\, e^{\beta(\mu_h k-\frac{U}{2} k(k-1))}}
{\sum_k  e^{\beta(\mu_h k-\frac{U}{2} k(k-1))}},
\end{equation}
The function $Q_k(U,\mu)$ appearing in Eq.~(\ref{E:matrixMSL}) is
exactly the same as defined in Eq.~(\ref{E:lobo}). 
Since $F_{h,h'}$ is a real and positive matrix, 
Perron-Frobenius theorem \cite{B:Meyer} ensures that its 
maximal eigenvalue $\phi_{\rm M}(\beta,U,\mu)$ is real and positive. 
Hence, the fixed point $\alpha_h=0$ is unstable --- and the system behaves like a superfluid ---
only when $t>t_c(\beta,U,\mu)=\phi_{\rm M}^{-1}$. 
Note that the  critical curve separating the superfluid and the normal
domains can also be defined as the lowest positive $t_c$ such that 
$P_F(t_c^{-1})=0$, where  $P_F(\lambda)$ is the characteristic polynomial
of matrix $F_{h,h'}$.

In the normal phase ($t<t_c$ and $\alpha_h=0$, $\forall~h$) the local density 
of particles at site $h$ is:
\begin{equation}
\label{E:rhoSL}
\rho_h(U,\mu,\beta)=\langle n_h \rangle = 
 \frac{\sum_k k e^{\beta((\mu_h k-\frac{U}{2} k(k-1))}}
{\sum_k  e^{\beta(\mu_h k-\frac{U}{2} k(k-1))}}
\end{equation}
Analogously to  the homogeneous case, for sufficiently low temperatures there
exist intervals of $\mu$ where the local particle density is arbitrarily close
to an integer value. In detail, introducing a small parameter $\epsilon$,
$|\rho_h(U,\mu,\beta)-k^*|<\epsilon$ if $\mu\in {\cal M}_h(k^*,\epsilon)$, where\[
{\cal M}_h(0,\epsilon)=\left]-\infty, v_h+\frac{\ln(\epsilon)}{\beta}\right[
\]
and, for any positive integer $k^*$,
\begin{equation}
\label{E:intv}
{\cal M}_h(k^*,\epsilon)\!=\!\left]U (k^*\!-\!1)\!+\!v_h\!-\!\frac{\ln(\epsilon)}{\beta},U k^*  \!+\!v_h\!+\!\frac{\ln(\epsilon)}{\beta}\right[.
\end{equation}
This in particular means that the average filling
\begin{equation}
\overline{\rho(U,\mu,\beta)}=\ell^{-1}\sum_{h=1}^{\ell} \rho_h(U,\mu,\beta)
\end{equation}
remains very close to the rational value 
\begin{equation}
\overline{k^*}=\ell^{-1}\sum_{h=1}^{\ell} k^*_h
\end{equation}
as long as the chemical potential belongs to the interval 
\begin{equation}
\label{E:inters}
{\cal M}(\{k^*_h\},\{v_h\},\epsilon)= \bigcap_{h=1}^{\ell} {\cal M}_h(k_h^*,\epsilon)
\end{equation}
where $\{k^*_h\}_{h=1}^\ell$ is a set of non-negative integers such that ${\cal M}\neq \emptyset$.
If, conversely, $\alpha_h=0$ but $\mu \not \in {\cal M}$, the average filling
is not close to a rational number and it significantly varies with varying
$\mu$. That is to say, the compressibility is significantly different from
zero, and the system behaves like a normal fluid, owing to thermal
fluctuations.

In the zero-temperature limit the normal phase behaviour disappears and a
(Mott)insulator-superfluid transition is recovered. Similar to the homogeneous
case, the $\mu/U$-$t/U$ zero-temperature phase diagram consists 
of a superfluid domain and a series of insulating 
Mott-lobes. However, the average filling and compressibility within these
lobes are exactly $\overline{k^*}$ and zero, respectively. 
Therefore, as it is expected \cite{CM:Santos}, we obtain that
superlattices can display an insulating-Mott behaviour even for some critical
rational fillings.
\begin{figure}
\begin{center}
\includegraphics[width=8.5 cm]{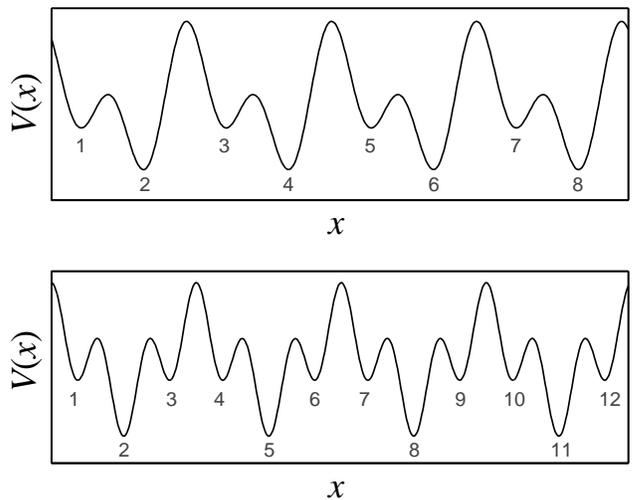}
\caption{
\label{F:SLpotper}
Schematic representation of the $\ell$-periodic optical superlattices
explicitly considered in Sec.~\ref{S:NHC}. The lattice sites correspond to the
local minima of the optical potential $V(x)$ . {\bf Upper panel}: $\ell=2$,
$v_j=v_{j+2} \neq v_{j+1}$, $\tau_{j,j+1} = \tau_{j+2,j+3} \neq
\tau_{j+1,j+2}$. {\bf Lower panel}:  $\ell=3$, $v_{j} = v_{j+2} \neq
v_{j+1}$, $\tau_{j,j+1} = \tau_{j+1,j+2}  \neq \tau_{j+2,j+3}$, $j=3 k+1$,
$k\in \mathbb{N}$.
}
\end{center}
\end{figure}

\begin{figure}
\begin{center}
\includegraphics[width=8.5 cm]{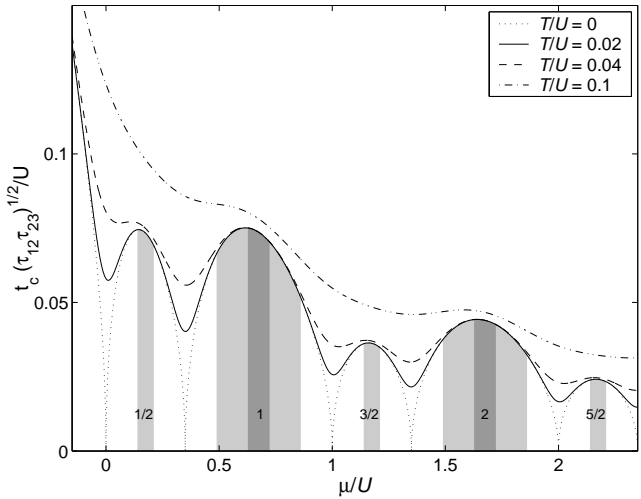}
\caption{\label{F:l2}
Phase diagram for the superlattice of periodicity $\ell=2$ with parameters
$v_{2 k}/U=0.35$, $v_{2 k+1}/U = 0$. Note that the possibly different values
of parameters $\tau_{h,h'}$ are conveniently absorbed
into $t_c$  by a
multiplicative rescaling. The filled areas correspond to the insulator-like
phase for $T=0.02$ (light gray) and $T=0.04$ (dark gray) as evaluated setting
$\epsilon=10^{-3}$ in Eq.~(\ref{E:inters}). The rational numbers denote the
particle density within the Mott-like domains.}
\end{center}
\end{figure}

Let us now analyze explicitly the two simple superlattices generated by the
trapping potential schematically represented in Fig.~\ref{F:SLpotper}.
 The upper panel of this figure corresponds to the simplest
choice, namely a superlattice of periodicity $\ell=2$. In this case the
maximal eigenvalue of matrix $F$ can be evaluated analytically, and the
resulting critical value of $t$ turns out to be
\begin{equation}
t_c(U,\mu,\beta)=\sqrt{\frac{1}{F_{1,2}F_{2,1}}}
\end{equation}
The ensuing mean-field phase-diagram for a particular choice of the parameters
is displayed in Fig.~\ref{F:l2}.
As mentioned above, rational (actually half-integer) filling Mott lobes
appear in the zero-temperature phase diagram (dotted curves). As the temperature increases, the regions
where the system is in a quasi insulating state shrink and eventually
disappear, according to Eq.~(\ref{E:inters}). It is interesting to observe that these regions may disappear at
different temperatures, depending on their filling. This is clearly shown in
Fig.~\ref{F:l2}, where the insulating regions relevant to $T=0.02$ (light
gray) and $T=0.04$ (dark gray) are shown. Note that in the latter case there
are only integer-filling (quasi) insulating regions.

\begin{figure}
\begin{center}
\includegraphics[width=8.5 cm]{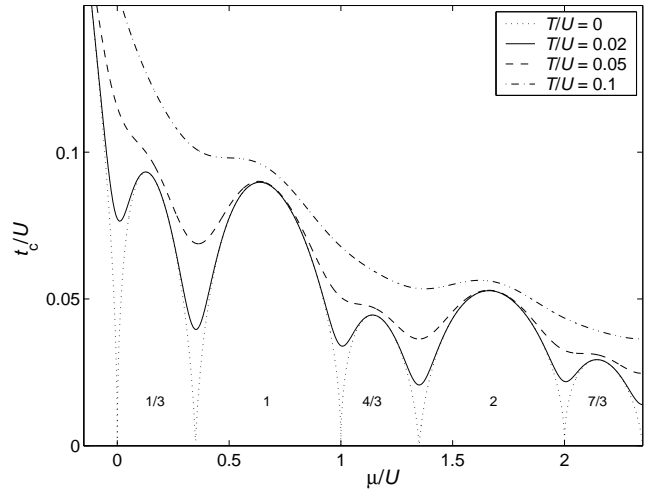}
\caption{\label{F:l3} Phase diagram for the superlattice of periodicity
  $\ell=3$ with parameters $v_{3 k}/U=v_{3 k+1}/U=0.35$, $v_{3 k+2}/U=0$,
  $\tau_{3 k,3k+1}=1/2$ and $\tau_{3 k+1,3k+2}=\tau_{3 k+2,3k+3}=1$. The rational numbers denote the particle density within the zero-temperature Mott insulator domains.}
\end{center}
\end{figure}

\begin{figure}
\begin{center}
\includegraphics[width=8.5 cm]{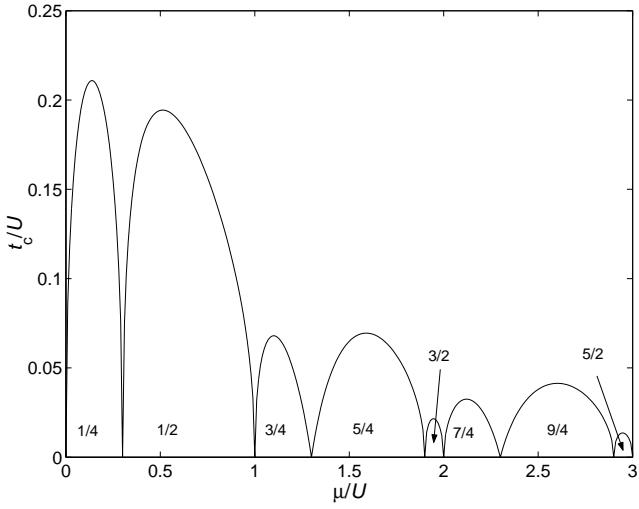}
\caption{
\label{F:l4} Zero-temperature phase diagram
for the superlattice of periodicity
  $\ell=4$ with parameters $v_{4 k}/U= 1.9$, $v_{4 k+2}/U= 0.3$, $v_{4 k+1}/U=1.3$, $v_{4 k+3}/U=0.0$,$\tau_{k,k+1}=1$. The rational numbers denote the
particle density within the Mott insulator domains. Note that, as we discussed
  in the text, for this particular choice of the well depths there are no integer
  critical fillings.}
\end{center}
\end{figure}

Matrix $F$ can be analytically diagonalized with a limited 
effort also for the special case of
3-periodic lattice illustrated in the lower panel of  Fig.~\ref{F:SLpotper},
where $v_1=v_3$ and $\tau_{1,2}=\tau_{2,3}$, so that 
$F_{1,2}=F_{3,2}$, $F_{1,3}=F_{3,1}$ and $F_{2,1}=F_{2,3}$. 
After a straightforward calculation one gets
\begin{equation}
t_c(U,\mu,\beta)=\frac{2}{F_{1,3}+\sqrt{F_{1,3}^2+8 F_{1,2}F_{2,1}}}
\end{equation}
The relevant phase diagram is shown in Fig.~\ref{F:l3}. 
Note that, similar to the previous example, the zero-temperature
Mott lobes can be divided into two classes depending on the relevant particle
density, which can be either  $k$ or $k+1/3$, where $k\in \mathbb{N}$. 
This simple behaviour is a consequence of our choice for the local potentials.
Of course, more structured choices result into a quite richer phase diagram,
where the particle density in the insulator-like regions is an integer 
multiple of $\ell^{-1}$. Note that some of these multiples are excluded 
if the energy offset between any two  sites $j$ and $h$ 
within the same supercell is an integer multiple of $U$. 
Indeed, in this situation, the intervals relevant to sites $j$ and $h$ defined by Eq.~(\ref{E:intv}) overlap exactly. Hence for some set of integers
the intersection defined by Eq.~(\ref{E:inters}) is empty.
This is exactly what happens in Fig.~\ref{F:l3}, where $v_1-v_3=0$.  
Interestingly, it is possible to devise
superlattices where  only fractional critical fillings are present, 
provided that
the energy offset between two lattice sites is larger than $U$. This is shown
in Fig~\ref{F:l4}, displaying the zero-temperature phase diagram for
a superlattice of periodicity $\ell=4$ as obtained by numerical
diagonalization of matrix $F$, Eq.~(\ref{F:l4}). 

\section{Conclusions}
\label{S:C}
In this paper we extend the mean-field approximation to the Bose-Hubbard
model introduced in Ref.~\cite{A:Sheshadri} to include thermal fluctuations.
An analytical study of the ensuing self-consistency equations allows us to 
determine the exact form of the boundary of the superfluid region at any
finite temperature. We also quantify the crossover from the normal fluid to
the insulator-like phase. 

Other than to the homogeneous $d$-dimensional lattice considered in the
original reference, we apply our method to a generic one-dimensional 
$\ell$-periodic superlattice, giving explicit results for $\ell=2$ and a
special case of $\ell=3$.
Results for more complex one-dimensional superlattices involve 
the evaluation of the maximal eigenvalue of a  $\ell\times\ell$ matrix.
Of course, this must be accomplished numerically even for relatively small
matrix sizes. Nevertheless this approach much less demanding and more precise 
than the (equivalent) fully numerical solution of the self-consistency equation. The latter actually
involves an iterative self-consistent diagonalization for each point of the
the mesh grid describing the phase diagram. 
Our technique can be extended also to generic $d$-dimensional superlattices, 
where the size of the matrix to be diagonalized is $s\times s$, $s$ being the
number of sites within a supercell.
We remark that the boundaries of the superfluid region as evaluated with our
method are valid also in the zero-temperature limit, thus providing the phase
diagram for the superfluid-insulator quantum transition. 
In particular this allows to find the analytic expressions of the (mean-field)
Mott lobes in the case explicitly considered above.

\acknowledgments
The work of P.B. has been entirely 
supported by MURST project {\it Quantum Information
  and Quantum Computation on Discrete Inhomogeneous Bosonic
  Systems}. A.V. also acknowledges partial financial support from the same
project. The authors wish to thank Vittorio Penna for fruitful discussion and comments.

\appendix
\section{}
\label{S:det}

In this section  the analytic expression for the boundary of the superfluid domain is derived in detail for the simple case of the homogeneous lattice. The generalization to the superlattice case is briefly discussed. 

As we mention above, in the homogeneous $d$-dimensional case, the self-consistency constraints, Eqs.~(\ref{E:sce}), reduce to a single equation in the variable $\alpha$, i.e. the site-independent superfluid parameter.
It proves useful to recast such equation in terms of the quantity $\gamma\equiv \alpha t$ as
\begin{equation}
\label{E:selfC}
\gamma = 
t \frac{{\rm Tr}\left(a \,e^{-\beta \bar {\cal H}}\right)}
{Z}= \frac{t}{4 d \beta} \frac{d}{d \gamma} \log(Z) \equiv f(\gamma) 
\end{equation}
where 
\begin{equation}
Z={\rm Tr}\left(e^{-\beta \bar {\cal H}}\right)
\end{equation}
is the grand-canonical partition function of the single site problem. The additional $1/2$ factor in Eq.~(\ref{E:selfC}) ensues from the equality $\alpha = \langle a\rangle  = \langle a^+\rangle \in {\mathbb R} $.

Let us now prove Eq.~(\ref{E:tc}) by discussing the stability character of the
fixed point $\gamma=0$ of the map~(\ref{E:selfC}). To this aim we truncate the on-site Fock basis considering states up to a
given number of particles  $n$ and denote  $\bar {\cal H}_n$ the relevant Hamiltonian matrix. The final result is obtained letting
  $n$ to infinity.

Introducing the set of eigenvalues of $\bar {\cal H}_n$, $\{E_k(\gamma)\}_{k=0}^{n}$
, Eq.~(\ref{E:selfC}) becomes
\begin{equation}
\label{E:selfCe}
\gamma =
 \frac{t}{4 d} \frac{\sum_{k=0}^{n} e^{-\beta \,E_k(\gamma)} \frac{d E_k(\gamma)}{d \gamma}}{\sum_{k=0}^n e^{-\beta \,E_k(\gamma)}} 
\end{equation}
Now, since $p_n(E_k(\gamma);\gamma)=0$, where $p_n(\lambda;\gamma)$ is the
characteristic polynomial of $\bar {\cal H}_n$
\begin{eqnarray}
0&=& \frac{d p_n(E_k(\gamma),\gamma)}{d \gamma} \nonumber
\\&=& \Big[ \frac{d E_k(\gamma)}{d  \gamma} \partial_\lambda p_n(\lambda,\gamma)  +  \partial_\gamma p_n(\lambda,\gamma) \Big]_{\lambda=E_k(\gamma)} 
\end{eqnarray}
so that it is possible to write
\begin{eqnarray}
\frac{d E_k(\gamma)}{d  \gamma} =-\left.\frac{\partial_\gamma p_n(\lambda,\gamma)}{\partial_\lambda p_n(\lambda,\gamma)}\right|_{\lambda=E_k(\gamma)} 
\end{eqnarray}
Denoting $p_n^{\{k\}}(\lambda;\gamma)$ the characteristic polynomial  of the
matrix obtained by discarding from $\bar{\cal H}_n$ the rows and columns labeled
by the set of indices $\{k\}$, and making use of the formula for the
derivative of a determinant \cite{B:Meyer}, one gets 
 \begin{eqnarray}
 \partial_\lambda p_n(\lambda,\gamma) &=& \sum_{k=0}^{n}
 p_n^{(k)}(\lambda;\gamma) \nonumber\\
 \partial_\gamma p_n(\lambda,\gamma) &=&\gamma 8\, d^2 \left[ \sum_{k=1}^{n} k
 \,  p_n^{(k,k-1)}(\lambda;\gamma)+P(\gamma)\right] \nonumber
 \end{eqnarray}
where the polynomial $P(\gamma)$ is homogeneous, so that $P(0)=0$. Hence
Eq.~(\ref{E:selfCe}) becomes
\begin{equation}
\label{E:selfCe2}
\gamma =
t\,\gamma \frac{\sum_{k=0}^{n} e^{-\beta \,E_k(\gamma)}
 q_n(E_k;\gamma) }{\sum_{k=0}^n e^{-\beta \,E_k(\gamma)}} \equiv f(\gamma)
\end{equation}
where
\begin{eqnarray}
q_n(\lambda;\gamma) &=& \frac{1}{4 d}\frac{\partial_\gamma p_n(\lambda,\gamma)}{\partial_\lambda
  p_n(\lambda,\gamma)} \nonumber \\
&=& 2d \frac{\sum_{h=1}^{n} h\,  p_n^{(h,h-1)}(\lambda;\gamma)+P(\gamma)}
{\sum_{h=0}^{n} p_n^{(h)}(\lambda;\gamma)}
\end{eqnarray}
According to standard treatment, $\gamma = 0$ is a stable solution of
Eq.~(\ref{E:selfCe2}) only if $|\partial_\gamma f(\gamma)|_{\gamma=0}<1$, i.e.
\begin{equation}
\label{E:critt}
t < \frac{\sum_{k=0}^n e^{-\beta \,E_k(0)}}{\sum_{k=0}^n q_n\left(E_k(0);0\right)e^{-\beta \,E_k(0)}}
\end{equation}
Observing that
\begin{equation}
p_n^{\{k\}}(\lambda;0) = \prod_{h\not\in\{k\}} \left(E_h(0)-\lambda\right)
\end{equation}
one gets
\begin{widetext}
\begin{equation}
q_n(E_k(0);0) = 2 d 
 \left[ k \,  p_n^{(k,k-1)}(E_k(0);0)+(k+1) \,  p_n^{(k+1,k)}(E_k(0);0) \right]
\left[p_n^{(k)}(E_k(0);0)\right]^{-1}
\end{equation}
\end{widetext}
where  we set $p_n^{(n,n+1)}(\lambda;\gamma)=p_n^{(-1,0)}(\lambda;\gamma)=0$.
Now, recalling that $E_k(0)=\frac{U}{2} k(k-1) -\mu k$ one gets
$ \lim_{n\to\infty} q_n(E_k(0);0) = Q_k(U,\mu)$, where the function $Q_k$
is defined in Eq.~(\ref{E:lobo}). Therefore, the limit $n\to \infty$ of
Eq.~(\ref{E:critt}) gives the desired result, Eq.~(\ref{E:tc}).

In the case of superlattices the parameters appearing in Hamiltonian (\ref{E:mfH1}) are periodic functions of the site labels. 
Here we focus on the one-dimensional $\ell$-periodic case, 
$v_j=v_{j+\ell}$ and
$\tau_{j,h}=(\delta_{h,j+1}+\delta_{h,j-1})\tau_{j+\ell,h+\ell}$.  
Since the superfluid parameters mirror the $\ell$-periodicity of the
superlattice, $\alpha_j=\alpha_{j+\ell}$, the
self-consistency conditions~(\ref{E:sce}) reduce to $\ell$ independent
equations.
Introducing the
parameters $\gamma_h\equiv t\alpha_h$ ($h=1,\dots,\ell$, $\gamma_{\ell+1}\equiv
\gamma_1$ and $\gamma_0\equiv \gamma_\ell$), equations~(\ref{E:sce}) can be 
recast as:
\begin{eqnarray}
\label{E:vmap}
\gamma_h &=& \frac{t}{4 \beta} \left[\frac{1}{\tau_{h,h-1}}\frac{d \log(Z_h)}{d \gamma_{h-1}} +
\frac{1}{\tau_{h,h+1}} \frac{d \log(Z_h)}{d \gamma_{h+1}} \right] \nonumber\\
& \equiv & f_h(\{\gamma_{h'}\})
\end{eqnarray}
where 
\begin{equation}
Z_h={\rm Tr}\left(e^{-\beta \tilde {\cal H}_h}\right)
\end{equation}
and
\begin{eqnarray}
\tilde{\cal H}_h & = &  
\frac{U}{2} n(n-1) - (\mu-v_h) n -\nonumber\\
&~& (\tau_{h,h-1}\gamma_{h-1}+\tau_{h,h+1} \gamma_{h+1}) (a+a^+)
\end{eqnarray}
A procedure similar to that detailedly illustrated in the case of homogeneous lattices allows to linearize Eq.~(\ref{E:vmap}), obtaining Eq.~(\ref{E:matmap}).


\begin{thebibliography}{10}

\bibitem{A:BAnderson}
B.~Anderson and M.~Kasevich, Science {\bf 282}, 1686 (1998).

\bibitem{A:Orzel}
C.~Orzel, A.~K. Tuchman, M.~Fenselau, M.~Yasuda and M.~A. Kasevich, Science
  {\bf 291}, 2386 (2001).

\bibitem{A:Morsch}
O.~Morsch, J.~H. Muller, M.~Cristiani, D.~Ciampini and E.~Arimondo, Phys. Rev.
  Lett. {\bf 87}, 140402 (2001).

\bibitem{A:Jaksch}
D.~Jaksch, C.~Bruder, J.~Cirac, C.~Gardiner and P.~Zoller, Phys. Rev. Lett.
  {\bf 81}, 3108 (1998).

\bibitem{A:Guidoni97}
L.~Guidoni, C.~Trich{\'{e}}, P.~Verkerk and G.~Grynberg, Phys. Rev. Lett. {\bf
  79}, 3363--3366 (1997).

\bibitem{A:Roth03}
R.~Roth and K.~Burnett, Phys. Rev. A {\bf 68}, 023604 (2003).

\bibitem{A:Blakie}
P.~B. Blakie and C.~W. Clark, J. Phys. B {\bf 37}, 1391 (2004).

\bibitem{CM:Santos}
L.~Santos, M.~Baranov, J.~Cirac, H.-U. Everts, H.~Fehrmann and M.~Lewenstein,
  e-print cond-mat/0401502 (2004).

\bibitem{A:Aizenman}
M.~Aizenman, E.~Lieb, R.~Seiringer, J.~Solovej and J.~Yngvason, e-print cond-mat/0403240 (2004).

\bibitem{A:Fisher}
M.~Fisher, P.~Weichman, G.~Grinstein and D.~S. Fisher, Phys. Rev. B {\bf 40},
  546 (1989).

\bibitem{N:Feshback}
The boson interaction $U$ can be also tuned experimentally making use of
  Feshback resonances.

\bibitem{A:FazioPR}
R.~Fazio and H.~van~der Zant, Phys. Rep. {\bf 355}, 235 (2001).

\bibitem{CM:Garcia}
J.~J. Garc{\'\i}a-Ripoll, M.~Martin-Delgado and J.~I. Cirac, e-print cond-mat/0404566 (2004).

\bibitem{A:Greiner}
M.~Greiner, I.~Bloch, O.~Mandel, T.~W. Hansch and T.~Esslinger, Phys. Rev.
  Lett. {\bf 87}, 160405 (2001).

\bibitem{A:Zwerger}
W.~Zwerger, J. Opt. B {\bf 5}, S9 (2003).

\bibitem{B:Sachdev}
S.~Sachdev, {\em Quantum Phase Transitions\/}, Cambridge University Press,
  (1999).

\bibitem{A:Sheshadri}
K.~Sheshadri, H.~Krishnamurthy, R.~Pandit and T.~Ramakrishnan, Europhys. Lett.
  {\bf 22}, 257 (1993).

\bibitem{A:Dicker}
D.~Dickerscheid, D.~van Oosten, P.~Denteneer and H.~Stoof, Phys. Rev. A {\bf
  68}, 043623 (2003).

\bibitem{A:Amico}
L.~Amico and V.~Penna, Phys. Rev. Lett. {\bf 80}, 2189 (1998).

\bibitem{A:vanOosten}
D.~van Oosten, P.~van~der Straten and H.~Stoof, Phys. Rev. A {\bf 63}, 053601
  (2001).

\bibitem{CM:Jain}
P.~Jain and C.~Gardiner, e-print cond-mat/0404642 (2004).

\bibitem{A:Freericks1}
J.~K. Freericks and M.~Monien, Europhys. Lett. {\bf 26}, 545--550 (1994).

\bibitem{CM:scpe}
P.~Buonsante, V.~Penna and A.~Vezzani, e-print cond-mat/0406467 (2004).

\bibitem{A:Elstner99a}
N.~Elstner and H.~Monien, Phys. Rev. B {\bf 59}, 12184 (1999).

\bibitem{A:Kuehner}
T.~K{\"{u}}hner and H.~Monien, Phys. Rev. B {\bf 58}, R14741 (1998).

\bibitem{A:Batrouni}
G.~Batrouni, R.~Scalettar and G.~Zimanyi, Phys. Rev. Lett. {\bf 65}, 1765
  (1990).

\bibitem{A:Kashurnikov96}
V.~Kashurnikov, A.~Krasavin and B.~Svistunov, JETP Lett. {\bf 64}, 99 (1996).

\bibitem{A:Kampf}
A.~Kampf and G.~Zimanyi, Phys. Rev. B {\bf 47}, 279 (1993).

\bibitem{A:Amico96}
L.~Amico, M.~Rasetti and R.~Zecchina, Physica A {\bf 230}, 300 (1996).

\bibitem{CM:Giampaolo}
S.~Giampaolo, F.~Illuminati, G.~Mazzarella and S.~D. Siena, e-print cond-mat/0403145 (2004).

\bibitem{CM:Plimak}
L.~Plimak, M.~Fleischhauer and M.~Olsen, e-print cond-mat/0309587 (2004).

\bibitem{N:real}
The superfluid parameters are real since the operators $a_j$ and $a_j^+$ have a
  real representation on the Fock basis.

\bibitem{A:Polkov}
A.~Polkovnikov, S.~Sachdev and S.~Girvin, Phys. Rev. A {\bf 66}, 053607 (2002).

\bibitem{A:BHComb}
P.~Buonsante, R.~Burioni, D.~Cassi, V.~Penna and A.~Vezzani, e-print cond-mat/0405520 (2004).

\bibitem{B:Meyer}
C.~Meyer, {\em Matrix Analysis and Applied Linear Algebra\/}, SIAM, (2001).

\end{thebibliography}

\end{document}